\documentclass[conference]{IEEEtran}
\IEEEoverridecommandlockouts
\usepackage{cite}
\usepackage{amsmath,amssymb,amsfonts}
\usepackage{algorithmic}
\usepackage{graphicx}
\usepackage{textcomp}
\usepackage{xcolor}
\usepackage{booktabs}
\usepackage{multirow}
\usepackage{url}
\usepackage{xurl}

\newcommand{\name}[1]{\def\paperauthors{#1}}
\newcommand{\address}[1]{\author{%
  \IEEEauthorblockN{\paperauthors}%
  \IEEEauthorblockA{#1}}}

\def\BibTeX{{\rm B\kern-.05em{\sc i\kern-.025em b}\kern-.08em
    T\kern-.1667em\lower.7ex\hbox{E}\kern-.125emX}}
\begin{document}

\title{Hearing Like Humans? Sound Symbolism and Perceptual Alignment in Speech Language Models
}

% \author{
% \IEEEauthorblockN{Yun-Shao Tsai\textsuperscript{*}\thanks{\textsuperscript{*}Equal Contribution.}}
% \IEEEauthorblockA{\textit{Graduate Institute of} \\
% \textit{Communication Engineering} \\
% \textit{National Taiwan University}\\
% Taipei, Taiwan \\
% r14942093@ntu.edu.tw}
% \and
% \IEEEauthorblockN{Chun-Wei Chen\textsuperscript{*}}
% \IEEEauthorblockA{\textit{Graduate Institute of} \\
% \textit{Electrical Engineering} \\
% \textit{National Taiwan University}\\
% Taipei, Taiwan \\
% r14921061@ntu.edu.tw}
% \and
% \IEEEauthorblockN{Chee-En Yu\textsuperscript{*}}
% \IEEEauthorblockA{\textit{Graduate Institute of} \\
% \textit{Electrical Engineering} \\
% \textit{National Taiwan University}\\
% Taipei, Taiwan \\
% r14921042@ntu.edu.tw
% }
% \and
% \IEEEauthorblockN{Yi-Cheng Lin\textsuperscript{*}}
% \IEEEauthorblockA{\textit{Graduate Institute of} \\
% \textit{Communication Engineering} \\
% \textit{National Taiwan University}\\
% Taipei, Taiwan \\
% f12942075@ntu.edu.tw}
% \and
% \IEEEauthorblockN{Hung-yi Lee}
% \IEEEauthorblockA{\textit{Artificial Intelligence Center of} \\
% \textit{Research Excellence} \\
% \textit{National Taiwan University}\\
% Taipei, Taiwan \\
% hungyilee@ntu.edu.tw}
% }
\name{Yun-Shao Tsai$^{1, *}$, 
Chun-Wei Chen$^{2, *}$, 
Chee-En Yu$^{2, *}$, 
Yi-Cheng Lin$^{1, *}$, 
Hung-yi Lee$^{1, 2, 3}$}
\address{$^{1}$ Graduate Institute of Communication Engineering, National Taiwan University, Taiwan \\
$^{2}$ Graduate Institute of Electrical Engineering, National Taiwan University, Taiwan \\
$^{3}$ NTU Artificial Intelligence Center of Research Excellence (NTU AI-CoRE), Taiwan \\
\{r14942093, r14921061, r14921042, f12942075, hungyilee\}@ntu.edu.tw \\
\thanks{\textsuperscript{*}Equal Contribution.}}

\maketitle

\begin{abstract}
Sound symbolism, the human tendency to map speech sounds to perceptual qualities such as roundness or sharpness, arises primarily from the acoustics of speech rather than spelling.
Whether Speech Language Models (SLMs) share this tendency remains open, as prior evaluations rely on text or images rather than real speech.
We study it using genuine human speech recordings, comparing model judgments against human data across the auditory, crossmodal, and visual components of the effect.
We find that SLMs' auditory judgments align poorly with human perception and miss the acoustic cues, such as spectral tilt, that drive human intuitions, and open-weight models cannot reliably link a heard sound to its corresponding shape.
With a visual-only control ruling out shape perception, the weakness localizes to how speech is represented, suggesting that perceptual alignment depends not on stronger vision but on speech representations that capture the cues humans hear.
\end{abstract}

\begin{IEEEkeywords}
Sound Symbolism, Speech Language Model, Crossmodal Correspondence, Perceptual Alignment
\end{IEEEkeywords}

\section{Introduction}
\label{sec:intro}
Sound symbolism is a phenomenon in which people share a common intuition about what a word ``feels like'' based solely on how it sounds, associating certain phonetic patterns with perceptual meanings~\cite{hinton1994sound,dingemanse2015arbitrariness}.
For example, in the bouba/kiki experiment, participants are presented with two pseudowords ``bouba'' and ``kiki'' alongside a rounded and a pointed shape; roughly 95\% of people match ``bouba'' to the rounded shape and ``kiki'' to the pointed one~\cite{kohler1967gestalt,ramachandran2001}, and the effect holds across diverse languages, cultures, and writing systems~\cite{cwiek2021bouba}.
The effect stems from the acoustic properties of speech sounds~\cite{ohala1984,sapir1929study}. Smooth, voiced, low-frequency sounds like (/b/, /u/) pair with rounded shapes, while abrupt, high-frequency sounds like (/k/, /i/) pair with jagged ones. The brain integrates these cues into a stable crossmodal correspondence~\cite{lacey2020stimulus}.
This raises a largely unexplored question: do Speech Language Models (SLMs) acquire analogous sound-shape intuitions from their training data?

This gap matters for human--machine alignment, with the increasing deployment of SLMs in creative applications like generating images from spoken descriptions~\cite{wang2021generating}, misaligned sound-symbolic intuitions can produce outputs that strike users as incongruent.
If a user requests a ``bomu'' robot and the model produces a sharp, angular form, the result contradicts the expectations that the name evokes for most listeners.
Understanding whether and how SLMs encode sound symbolism is thus a prerequisite for building systems whose outputs are perceptually aligned with human sensory associations.

This alignment is non-trivial.
SLMs are trained for tasks such as speech recognition and audio understanding~\cite{NEURIPS2025_3babb6b4,wu2025step,ding2025kimi,cui2026minicpm,yang-etal-2025-towards-holistic,xu2025qwen3}.
Even when their training data spans multiple modalities, they are never supervised to map low-level acoustic features to perceptual qualities such as roundness or sharpness.
Any sound-symbolic behavior they exhibit would therefore emerge implicitly from their training data and objectives, rather than from direct supervision, and whether such intuitions arise as a byproduct of learning speech is an open empirical question.
Among the many forms of sound symbolism, we focus on the bouba/kiki effect because it is among the most robust: it holds across many languages, writing systems, and ages~\cite{cwiek2021bouba}, making it one of the clearest cases of a shared perceptual mapping.
This consistency provides a well-defined human baseline against which model behavior can be measured.

\begin{figure}
    \centering
    \includegraphics[width=0.5\textwidth]{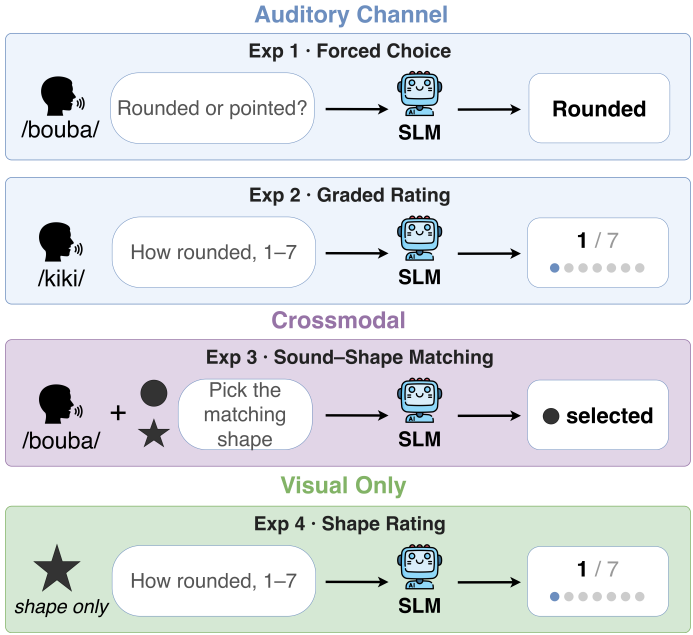}
    \vspace{-5mm}
    \caption{Experimental design for probing the bouba/kiki effect in SLMs. Experiments 1–2 isolate the auditory channel via forced-choice and graded ratings of spoken pseudowords; Experiment 3 tests crossmodal sound-to-shape matching; Experiment 4 is a visual-only ablation.}
    \label{fig:overview} 
    \vspace{-7mm}
\end{figure}

Previous works are confined to text and image modalities~\cite{alper2023kiki,zhao2026evidence}, and existing work in the audio domain relies on annotations synthetically generated by Large Language Models rather than human perceptual data~\cite{jeong2026language}.
Thus, it remains unknown whether SLMs exhibit the effect in speech modality where it originates, and whether their behavior matches that of humans.
We formulate four research questions that move from the auditory channel to crossmodal matching:
\begin{itemize}
    \item \textbf{RQ 1.} Do SLMs judge speech sounds as round or pointed the way humans do?
    \item \textbf{RQ 2.} Are SLMs' sound-symbolic judgments driven by the same acoustic cues that drive human judgments, such as spectral tilt?
    \item \textbf{RQ 3.} Can SLMs match a heard sound to the corresponding shape?
    \item \textbf{RQ 4.} If matching fails, can the failure be attributed to the auditory side rather than the visual one?
\end{itemize}

Figure~\ref{fig:overview} summarizes our experiments.
The results reveal that on the auditory side, the models' sound-symbolic judgments diverge from human perception: the best model reaches barely half of the human ceiling, and representational similarity analysis (RSA) shows that models do not exploit the acoustic cues that drive human judgments.
Crucially, the open-weight models fail to connect these auditory representations to the visual domain in the way humans do, while a frontier model succeeds at the task despite similarly weak auditory alignment.
On the visual side, by contrast, omni models (those that accept both audio and image inputs) closely agree with human shape ratings, demonstrating that they can represent the rounded-pointed distinction with high fidelity when the input is visual rather than acoustic.
Additionally, we investigate which layers form these perceptual judgments.
Applying the logit lens~\cite{logit_lens} to the strongest open-weight model, we find that for both image and audio inputs, the perceptual decision forms only in the deepest layers.
Our contributions are as follows:
\begin{enumerate}
    \item We conduct the first systematic evaluation of sound symbolism in SLMs using human-recorded speech and validating model behavior against human perceptual data, testing eight models in 25 languages (ten analyzed) and materials from established psycholinguistic studies.\footnote{Code:\url{https://github.com/danielqwer/Sound-Symbolism-SLM}}
    \item We decompose the bouba/kiki effect into auditory, visual, and crossmodal components, revealing that open-weight SLMs fail specifically at the auditory-visual integration that is central to human sound symbolism.
    \item We localize this failure using representational similarity analysis and logit-lens analysis, showing that perceptual decisions form only in the deepest layers.
\end{enumerate}

\section{Related Work}
\label{sec:related}
\subsection{Prior Benchmarks in Speech Language Models}
To assess the evolving capabilities of SLMs, prior work has introduced diverse benchmarks across various auditory and linguistic dimensions.
Initial evaluations focus mainly on speech understanding and general acoustic comprehension \cite{dynamicsuperb2, chen2026voicebench, speechifeval, listentthinkunderstand, wang-etal-2025-audiobench, sakshi2024mmaumassivemultitaskaudio, mmaupro}.
Beyond basic understanding, studies have also explored paralinguistic phenomena and acoustic conditions related to human concepts \cite{dynamicsuperb1,zhou2025echomind}.
However, while these benchmarks probe explicit acoustic, environmental, and paralinguistic cues, whether SLMs' capabilities align with the more implicit patterns of human auditory cognition, particularly sound symbolism, remains largely uninvestigated.

\subsection{Auditory Cognition in Speech Language Models}
A parallel line of work has begun to probe sound symbolism in multimodal models.
Vision--language models have been shown to exhibit the bouba/kiki effect through text--image associations~\cite{alper2023kiki, iida2024investigating, shinto2024analyzing}, and Loakman et al.~\cite{loakman2024} found that magnitude symbolism in multimodal LLMs scales with model size.
Despite these conceptual advances, current evaluations are mostly limited to text and vision, leaving actual audio processing largely unexplored.
Furthermore, recent work in the audio domain~\cite{jeong2026language} uses mostly synthetically generated labels and natural dictionary words, which can introduce lexical memorization effects.
To bridge this gap and isolate the acoustic signal itself, we bypass textual proxies and evaluate models directly against human perceptual data, grounding our evaluation in established psycholinguistic pseudowords, raw audio waveforms, and empirically validated human ratings.

\section{Methodology}
\label{sec:methodology}
We answer the four questions of Section~\ref{sec:intro} with four experiments.
Experiments~1 and~2 stay in the auditory channel and ask whether the model hears words as rounded or pointed, as people do (RQ1), through a forced choice and a graded rating.
The graded ratings also allow us to ask which acoustic cues the model's judgments track (RQ2).
Experiment~3 then brings in vision and asks whether the model can match a heard sound to the corresponding shape (RQ3).
A matching failure could originate on the auditory or the visual side, so Experiment~4 removes the audio and has the model rate the shapes alone (RQ4).
If the model rates the shapes as people do, its visual side is intact, and any matching failure can then be attributed to the auditory side rather than to shape perception.

\subsection{Experiment 1 (choosing rounded or pointed for a sound)}
\label{sec:exp1}
\noindent\textbf{Stimuli.}
Experiment~1 builds on the cross-linguistic study of \'{C}wiek et al.~\cite{cwiek2021bouba}, who tested the bouba/kiki effect in 25 languages.
In that study, listeners heard two pseudowords, \emph{bouba} and \emph{kiki}, and matched each to either a rounded or a pointed shape.
We reuse the original recordings of both words, both spoken by a single phonetician, so the model hears exactly the same sounds the human participants did.
We run the task in the same 25 languages, with one prompt translated into each.
Because the audio is fixed and the prompt varies only in language, any difference we see across languages reflects the prompt language rather than the sound.
 
\noindent\textbf{Task.}
The model hears one recording and chooses whether it sounds round or pointed.
The two sounds, 25 prompt languages, and 30 repetitions per item give 1{,}500 trials per model.
Since no shape is shown, this is a classification task rather than the sound-to-shape matching that human participants performed, a difference we keep in mind when comparing the results with the human rates.

\noindent\textbf{Analysis.}
Among the 25 languages, we keep the ten with the lowest unparseable-answer rates; in each retained language, the unparseable rate stays below 10\% for every model, whereas the excluded languages yield too few parseable responses for a trustworthy rate.
For each language, we report how often the model chooses rounded for \emph{bouba} and pointed for \emph{kiki}, the direction predicted by the bouba/kiki effect, and test each rate against chance with a two-sided binomial test.
To summarize the effect across languages, we compare the \emph{bouba} and \emph{kiki} rates over the ten languages with a Wilcoxon signed-rank test.

\subsection{Experiment 2 (rating a sound for rounded and pointed)}
\label{sec:exp2}
 
\noindent\textbf{Stimuli.}
Experiment~2 uses 536 of the 537 pseudowords from McCormick et al.~\cite{mccormick2015}.
We exclude one word because it differs from its counterpart in Lacey et al.~\cite{lacey2020stimulus}, whose dissimilarity matrix our second-order RSA requires.
All were recorded by one female American-English speaker with flat intonation.
The consonants cover sonorants, fricatives, affricates, and stops, with voiced and voiceless examples, and the vowels vary in frontness and rounding.
Because the sounds vary continuously along the acoustic dimensions that people use for sound symbolism, we use graded rating rather than a binary choice.
 
\noindent\textbf{Task.}
In two separate prompts, the model hears the pseudoword and rates it on a 7-point scale, once for rounded and once for pointed.
We use two scales because human rounded and pointed ratings are only weakly negatively correlated for pseudowords, unlike the strong negative correlation observed for shapes.
A sound can strike a listener as somewhat rounded and somewhat pointed at once, and two scales let us see whether a model does the same.
 
\noindent\textbf{Analysis.}
For each word, we combine its mean rounded and mean pointed into one rounded-to-pointed score, $(\text{rounded} + (8 - \text{pointed}))/2$, where higher values mean rounder.
We then run three analyses: a per-word correlation with human ratings, a comparison of dissimilarity structures, and an acoustic-cue analysis.
First, to test whether the model orders the words from rounded to pointed as people do, we correlate the 536 model scores with human means using Pearson $r$ and Spearman $r_s$, and assess the statistical significance of the resulting correlation coefficients.
We read each correlation against a human ceiling: the split-half reliability of the human ratings with a Spearman--Brown correction~\cite{spearman1910correlation, brown1910some}, computed separately for $r$ and $r_s$.
This reliability is the highest correlation a model could reach, since no model can match the human mean more closely than the two halves match each other.
Second, to compare how the model and people place the words relative to one another, we build a $536\times536$ dissimilarity matrix for each side and take the Spearman rank correlation ($\rho$) between them, with significance from a Mantel permutation test~\cite{mantel1967detection} (10{,}000 permutations).
The human matrix comes from Lacey et al.~\cite{lacey2020stimulus}: each word's ratings are stacked into one vector, one entry per rater, and the dissimilarity between two words is one minus the Pearson correlation between their vectors, so that two rounded or two pointed words come out similar while a rounded and a pointed word come out strongly dissimilar.
The model gives one score per word, so its matrix uses the absolute difference between scores instead; comparing the matrices with a rank correlation absorbs this difference in construction.
Third, to ask which acoustic cues underlie the model's judgments and whether they are the cues people rely on, such as spectral tilt, we correlate the model's matrix with the ten acoustic-parameter matrices of Lacey et al.~\cite{lacey2020stimulus}, which span spectral, temporal, and voice-quality measures.
We assess each correlation with the same Mantel permutation test and compare the resulting cue profile with the human one.

\subsection{Experiment 3 (matching a sound to a shape)}
\label{sec:exp3}

\noindent\textbf{Stimuli.}
Experiment~3 pairs the two recordings from Experiment~1 with two shapes from '{C}wiek et al.~\cite{cwiek2021bouba}: one rounded and one pointed.
On every trial the model hears one recording and chooses between these two shapes, so that any variation across trials reflects the sound rather than the shapes.

\noindent\textbf{Task.}
The model picks the shape that fits the sound; of our four experiments, this is the closest to the task the human participants performed in \'{C}wiek et al.~\cite{cwiek2021bouba}.
As in Experiment~1, the prompt is given in each of the 25 languages.

\noindent\textbf{Analysis.}
As in Experiment~1, we keep the ten languages with the lowest unparseable-answer rates; because these rates are computed per experiment, the retained set can differ from that of Experiment~1.
Since our focus is auditory rather than cross-linguistic, we report matching rates averaged across ten languages rather than per language, referencing human matching rates from \'{C}wiek et al.~\cite{cwiek2021bouba} on the same ten languages.

\subsection{Experiment 4 (rating a shape from rounded to pointed)}
\label{sec:exp4}
 
\noindent\textbf{Stimuli.}
Experiment~4 uses the 90-shape set of Lacey et al.~\cite{lacey2020stimulus}.
For the human comparison we use people's ratings of these shapes, also from Lacey et al.~\cite{lacey2020stimulus}.
 
\noindent\textbf{Task.}
Experiment~4 is the visual-only ablation of Experiment~3, with the audio removed so the model judges the shapes alone.
Instead of matching, the model sees one shape with no sound and rates it on the two scales of Experiment~2.
The ratings show whether a model can judge shapes like people even without sound.
 
\noindent\textbf{Analysis.}
We assess the model's shape ratings against people's with the first two analyses of Experiment~2, now over the 90 shapes.
We correlate the model and human means (Pearson $r$ and Spearman $r_s$), and we compare their $90\times90$ dissimilarity matrices (Spearman $\rho$), built as in Experiment~2.
If the model matches people on both, its visual judgment is intact, which lets us attribute the weak matching of Experiment~3 to the auditory side rather than to shape perception.

\begin{table*}[t]
\centering
\caption{Experiment~1. For each language, the table reports how often the model labels \emph{bouba} as rounded (\textbf{B}) and \emph{kiki} as pointed (\textbf{K}). In each column the model closest to the human rate is shown in \textbf{bold}. A star ($^{*}$) marks a rate significantly different from chance by a two-sided binomial test against $0.5$ ($p<.05$). The \textbf{B\,vs\,K} column gives the direction of the bouba/kiki difference pooled across the ten languages; a dagger ($^{\dagger}$) marks a significant Wilcoxon signed-rank test over the ten languages ($p<.05$). The evaluated models are divided into audio-only models (top) and omni models (bottom).}
\vspace{-2mm}
\label{tab:exp1}
\setlength{\tabcolsep}{3pt}
\resizebox{\textwidth}{!}{%
\begin{tabular}{l*{20}{r}c}
\toprule
\multirow{2}{*}{Model} & \multicolumn{2}{c}{en} & \multicolumn{2}{c}{ja} & \multicolumn{2}{c}{zh-cn} & \multicolumn{2}{c}{pt} & \multicolumn{2}{c}{fr} & \multicolumn{2}{c}{tr} & \multicolumn{2}{c}{it} & \multicolumn{2}{c}{de} & \multicolumn{2}{c}{pl} & \multicolumn{2}{c}{hu} & \multirow{2}{*}{B\,vs\,K} \\
\cmidrule(lr){2-3}\cmidrule(lr){4-5}\cmidrule(lr){6-7}\cmidrule(lr){8-9}\cmidrule(lr){10-11}\cmidrule(lr){12-13}\cmidrule(lr){14-15}\cmidrule(lr){16-17}\cmidrule(lr){18-19}\cmidrule(lr){20-21}
 & \multicolumn{1}{c}{B} & \multicolumn{1}{c}{K} & \multicolumn{1}{c}{B} & \multicolumn{1}{c}{K} & \multicolumn{1}{c}{B} & \multicolumn{1}{c}{K} & \multicolumn{1}{c}{B} & \multicolumn{1}{c}{K} & \multicolumn{1}{c}{B} & \multicolumn{1}{c}{K} & \multicolumn{1}{c}{B} & \multicolumn{1}{c}{K} & \multicolumn{1}{c}{B} & \multicolumn{1}{c}{K} & \multicolumn{1}{c}{B} & \multicolumn{1}{c}{K} & \multicolumn{1}{c}{B} & \multicolumn{1}{c}{K} & \multicolumn{1}{c}{B} & \multicolumn{1}{c}{K} & \\
\midrule
Human & 0.88\phantom{$^{*}$} & 0.88\phantom{$^{*}$} & 0.89\phantom{$^{*}$} & 0.95\phantom{$^{*}$} & 0.52\phantom{$^{*}$} & 0.54\phantom{$^{*}$} & 0.67\phantom{$^{*}$} & 0.62\phantom{$^{*}$} & 0.92\phantom{$^{*}$} & 0.78\phantom{$^{*}$} & 0.61\phantom{$^{*}$} & 0.45\phantom{$^{*}$} & 0.88\phantom{$^{*}$} & 0.76\phantom{$^{*}$} & 0.91\phantom{$^{*}$} & 0.79\phantom{$^{*}$} & 0.89\phantom{$^{*}$} & 0.81\phantom{$^{*}$} & 0.95\phantom{$^{*}$} & 0.75\phantom{$^{*}$} & -- \\
\midrule
% Audio-only models (top)
Audio Flamingo 3 & 0.53\phantom{$^{*}$} & 0.57\phantom{$^{*}$} & 0.67$^{*}$ & 0.23$^{*}$ & 0.97$^{*}$ & 0.20$^{*}$ & \textbf{0.77}$^{*}$ & 0.27$^{*}$ & \textbf{0.87}$^{*}$ & 0.21$^{*}$ & 0.83$^{*}$ & 0.17$^{*}$ & 0.00$^{*}$ & 1.00$^{*}$ & 0.14$^{*}$ & \textbf{0.78}$^{*}$ & 0.21$^{*}$ & 0.75$^{*}$ & 0.30$^{*}$ & 0.48\phantom{$^{*}$} & b$>$k\phantom{$^{\dagger}$} \\
Kimi-Audio & \textbf{0.90}$^{*}$ & 0.07$^{*}$ & 0.97$^{*}$ & 1.00$^{*}$ & 1.00$^{*}$ & 0.24$^{*}$ & 0.97$^{*}$ & 0.10$^{*}$ & \textbf{0.87}$^{*}$ & 0.47\phantom{$^{*}$} & 0.64\phantom{$^{*}$} & 0.86$^{*}$ & 0.67\phantom{$^{*}$} & 0.43\phantom{$^{*}$} & \textbf{1.00}$^{*}$ & 0.53\phantom{$^{*}$} & 1.00$^{*}$ & 0.03$^{*}$ & 0.33\phantom{$^{*}$} & 0.36\phantom{$^{*}$} & b$>$k$^{\dagger}$ \\
Step-Audio2 & 0.20$^{*}$ & \textbf{0.77}$^{*}$ & 0.00$^{*}$ & \textbf{0.97}$^{*}$ & 0.30$^{*}$ & 0.60\phantom{$^{*}$} & 0.80$^{*}$ & 0.17$^{*}$ & 0.80$^{*}$ & 0.20$^{*}$ & 0.73$^{*}$ & 0.27$^{*}$ & 0.47\phantom{$^{*}$} & \textbf{0.63}\phantom{$^{*}$} & 0.17$^{*}$ & 0.90$^{*}$ & 0.23$^{*}$ & \textbf{0.83}$^{*}$ & 0.50\phantom{$^{*}$} & 0.53\phantom{$^{*}$} & k$>$b\phantom{$^{\dagger}$} \\
GPT-Audio-1.5 & 1.00$^{*}$ & 1.00$^{*}$ & 1.00$^{*}$ & 1.00$^{*}$ & 1.00$^{*}$ & 1.00$^{*}$ & 1.00$^{*}$ & 1.00$^{*}$ & \textbf{0.97}$^{*}$ & \textbf{0.79}\phantom{$^{*}$} & 0.53\phantom{$^{*}$} & 1.00$^{*}$ & 1.00$^{*}$ & 1.00$^{*}$ & \textbf{1.00}$^{*}$ & 1.00$^{*}$ & 1.00$^{*}$ & 1.00$^{*}$ & \textbf{1.00}$^{*}$ & 1.00$^{*}$ & k$>$b\phantom{$^{\dagger}$} \\
\midrule
% Omni models (bottom)
Gemma4-E4B & 0.77$^{*}$ & 0.40\phantom{$^{*}$} & 0.77$^{*}$ & 0.50\phantom{$^{*}$} & 0.90$^{*}$ & \textbf{0.53}\phantom{$^{*}$} & 0.93$^{*}$ & \textbf{0.40}\phantom{$^{*}$} & 1.00$^{*}$ & 0.00$^{*}$ & 0.93$^{*}$ & 0.13$^{*}$ & \textbf{0.97}$^{*}$ & 0.07$^{*}$ & 0.73$^{*}$ & 0.20$^{*}$ & 0.97$^{*}$ & 0.23$^{*}$ & 0.77$^{*}$ & 0.00$^{*}$ & b$>$k$^{\dagger}$ \\
MiniCPM-o-4.5 & 0.97$^{*}$ & 0.00$^{*}$ & \textbf{0.87}$^{*}$ & 0.20$^{*}$ & 1.00$^{*}$ & 0.00$^{*}$ & 0.87$^{*}$ & 0.13$^{*}$ & 1.00$^{*}$ & 0.10$^{*}$ & 0.93$^{*}$ & 0.17$^{*}$ & \textbf{0.97}$^{*}$ & 0.00$^{*}$ & 0.80$^{*}$ & 0.40\phantom{$^{*}$} & \textbf{0.87}$^{*}$ & 0.20$^{*}$ & 0.70\phantom{$^{*}$} & 0.40\phantom{$^{*}$} & b$>$k$^{\dagger}$ \\
Qwen3-Omni & 0.33\phantom{$^{*}$} & 0.37\phantom{$^{*}$} & 0.77$^{*}$ & 0.17$^{*}$ & \textbf{0.70}$^{*}$ & \textbf{0.53}\phantom{$^{*}$} & 1.00$^{*}$ & 0.10$^{*}$ & 1.00$^{*}$ & 0.20$^{*}$ & \textbf{0.63}\phantom{$^{*}$} & \textbf{0.50}\phantom{$^{*}$} & 1.00$^{*}$ & 0.00$^{*}$ & 0.80$^{*}$ & 0.73$^{*}$ & 0.50\phantom{$^{*}$} & 0.50\phantom{$^{*}$} & 0.40\phantom{$^{*}$} & \textbf{0.63}\phantom{$^{*}$} & b$>$k\phantom{$^{\dagger}$} \\
Gemini3.5-Flash & 1.00$^{*}$ & 0.00$^{*}$ & 1.00$^{*}$ & 1.00$^{*}$ & 1.00$^{*}$ & 0.63\phantom{$^{*}$} & 1.00$^{*}$ & 1.00$^{*}$ & 1.00$^{*}$ & 1.00$^{*}$ & 1.00$^{*}$ & 1.00$^{*}$ & 1.00$^{*}$ & 0.10$^{*}$ & \textbf{1.00}$^{*}$ & 0.50\phantom{$^{*}$} & 1.00$^{*}$ & 0.50\phantom{$^{*}$} & \textbf{1.00}$^{*}$ & 1.00$^{*}$ & b$>$k\phantom{$^{\dagger}$} \\
\bottomrule
\end{tabular}%
}
\end{table*}

\subsection{Logit Lens Analysis}
\label{sec:logitlens}
The logit lens~\cite{logit_lens} traces how a model's prediction evolves across layers, revealing at which depth the internal representation first becomes aligned with the final output.
It does so by reading the residual stream at a given token position at each layer, projecting it through the model's final unembedding matrix, and applying a softmax to obtain a probability distribution over the output vocabulary.
This yields one distribution per layer: when it is diffuse, the intermediate representation does not yet point toward a specific output; when it concentrates on a single token, the representation has become decodable as a concrete prediction.
We apply this to the strongest open-weight model, reading the distribution over the digit tokens (1--7) at the final token position, where the hidden state has processed the entire input, for both audio and image stimuli.

\section{Experiment}
\label{sec:experiment}
\subsection{Setup}
\label{sec:setup}
 
\noindent\textbf{Models.}
The eight SLMs we evaluate fall into two groups: four \emph{omni} models that accept both audio and images, namely Qwen3-Omni~\cite{xu2025qwen3}, MiniCPM-o-4.5~\cite{cui2026minicpm}, Gemma4-E4B~\cite{google2025gemma4}, and Gemini3.5-Flash, which run all four experiments; and four \emph{audio-only} models, namely Step-Audio2~\cite{wu2025step}, Kimi-Audio~\cite{ding2025kimi}, Audio Flamingo 3~\cite{NEURIPS2025_3babb6b4}, and GPT-Audio-1.5, which run only the two auditory experiments (Experiments~1 and~2), since Experiments~3 and~4 require image input that these models cannot accept.
 
\noindent\textbf{Protocol.}
We sample 30 responses per item at temperature 0.7, so each score reflects the model's response distribution rather than one sample.
An item is one stimulus in one prompt language: a spoken word in Experiments~1 and~3, a pseudoword in Experiment~2, and a shape in Experiment~4.
In Experiments~1 and~3 we counterbalance option order by presenting each of the two options first on half of the 30 repeats, which distinguishes genuine choices from a bias toward the option listed first.
Models answer in free form, and an LLM judge (Qwen3.6-35B-A3B) extracts the chosen label from each reply; the judge only extracts labels and does not score models, so it introduces no quality judgment of its own.
To verify this extraction, we manually annotated a random sample of 500 replies per experiment; the judge's labels matched human annotation on all of them.
A reply is parseable when it commits to one of the response options and unparseable when it does not, for instance when the model refuses, answers off task, or settles on neither option; we compute each rate over the parseable trials only.
 
\begin{table}[t]
\centering
\caption{Experiment~2 (graded rating of pseudowords). Model agreement with humans on per-word scores (first-order) and on the $536\times536$ dissimilarity matrix (second-order RSA $\rho$). An asterisk ($^{*}$) marks correlations significant at $p<.05$. The evaluated models are divided into audio-only models (top) and omni models (bottom).}
\vspace{-2mm}
\label{tab:exp2}
\setlength{\tabcolsep}{6pt}
\begin{tabular}{lccc}
\toprule
\multirow{2}{*}{Model} & \multicolumn{2}{c}{First-order (per-word)} & Second-order \\
\cmidrule(lr){2-3}\cmidrule(lr){4-4}
 & Pearson $r$ & Spearman $r_s$ & RSA $\rho$ \\
\midrule
Human ceiling & \phantom{-}.843\phantom{$^{*}$} & \phantom{-}.838\phantom{$^{*}$} & \phantom{-}.377\phantom{$^{*}$} \\
\midrule
% Audio-only models (top)
Audio Flamingo 3 & -.033\phantom{$^{*}$} & -.014\phantom{$^{*}$} & \phantom{-}.008\phantom{$^{*}$} \\
Kimi-Audio & \phantom{-}.415$^{*}$ & \phantom{-}.421$^{*}$ & \phantom{-}.056$^{*}$ \\
Step-Audio2 & -.020\phantom{$^{*}$} & -.032\phantom{$^{*}$} & \phantom{-}.020$^{*}$ \\
GPT-Audio-1.5 & \phantom{-}.231$^{*}$ & \phantom{-}.200$^{*}$ & \phantom{-}.072$^{*}$ \\
\midrule
% Omni models (bottom)
Gemma4-E4B & \phantom{-}.044\phantom{$^{*}$} & \phantom{-}.054\phantom{$^{*}$} & \phantom{-}.004\phantom{$^{*}$} \\
MiniCPM-o-4.5 & \phantom{-}.283$^{*}$ & \phantom{-}.251$^{*}$ & \phantom{-}.063$^{*}$ \\
Qwen3-Omni & \phantom{-}.470$^{*}$ & \phantom{-}.461$^{*}$ & \phantom{-}.104$^{*}$ \\
Gemini3.5-Flash & \phantom{-}.392$^{*}$ & \phantom{-}.337$^{*}$ & \phantom{-}.019$^{*}$ \\
\bottomrule
\end{tabular}
\end{table}
\begin{table*}[t]
\centering
\caption{Experiment~2 acoustic-cue RSA. Spearman $\rho$ between each model's audio matrix and ten human acoustic-parameter matrices; \textbf{bold} marks each row's strongest cue. Asterisks mark $p<.05$.The evaluated models are divided into audio-only models (top) and omni models (bottom).}
\vspace{-2mm}
\label{tab:exp2_acoustic}
\setlength{\tabcolsep}{4pt}
\footnotesize
\begin{tabular}{l ccc ccccccc}
\toprule
\multirow{2}{*}{Model} & \multicolumn{3}{c}{Spectral / temporal} & \multicolumn{7}{c}{Voice quality} \\
\cmidrule(lr){2-4}\cmidrule(lr){5-11}
 & \shortstack{Spectral\\tilt} & \shortstack{Temporal\\modulation} & \shortstack{Speech\\envelope} & \shortstack{Auto-\\correlation} & HNR & \shortstack{Fraction\\unvoiced} & \shortstack{Pulse\\number} & Jitter & Shimmer & \shortstack{Pitch\\SD} \\
\midrule
Human & \phantom{-}\textbf{.431}\phantom{$^{*}$} & \phantom{-}.229\phantom{$^{*}$} & \phantom{-}.130\phantom{$^{*}$} & \phantom{-}.074\phantom{$^{*}$} & \phantom{-}.136\phantom{$^{*}$} & \phantom{-}.082\phantom{$^{*}$} & \phantom{-}.101\phantom{$^{*}$} & \phantom{-}.024\phantom{$^{*}$} & \phantom{-}.059\phantom{$^{*}$} & -.003\phantom{$^{*}$} \\
\midrule
% Audio-only models (top)
Audio Flamingo 3 & \phantom{-}.002\phantom{$^{*}$} & \phantom{-}.000\phantom{$^{*}$} & -.001\phantom{$^{*}$} & -.002\phantom{$^{*}$} & -.004\phantom{$^{*}$} & \phantom{-}.000\phantom{$^{*}$} & -.003\phantom{$^{*}$} & -.012\phantom{$^{*}$} & -.014\phantom{$^{*}$} & \phantom{-}\textbf{.020}\phantom{$^{*}$} \\
Kimi-Audio & \phantom{-}.018\phantom{$^{*}$} & \phantom{-}.025$^{*}$ & \phantom{-}.006\phantom{$^{*}$} & \phantom{-}.036$^{*}$ & \phantom{-}\textbf{.044}$^{*}$ & \phantom{-}.022$^{*}$ & \phantom{-}.041$^{*}$ & \phantom{-}.027$^{*}$ & \phantom{-}.006\phantom{$^{*}$} & \phantom{-}.034$^{*}$ \\
Step-Audio2 & \phantom{-}.026$^{*}$ & \phantom{-}\textbf{.027}$^{*}$ & \phantom{-}.023\phantom{$^{*}$} & -.010\phantom{$^{*}$} & -.025\phantom{$^{*}$} & \phantom{-}.000\phantom{$^{*}$} & -.017\phantom{$^{*}$} & -.006\phantom{$^{*}$} & -.011\phantom{$^{*}$} & -.011\phantom{$^{*}$} \\
GPT-Audio-1.5 & \phantom{-}.054$^{*}$ & \phantom{-}\textbf{.087}$^{*}$ & \phantom{-}.077$^{*}$ & \phantom{-}.033$^{*}$ & \phantom{-}.066$^{*}$ & \phantom{-}.011\phantom{$^{*}$} & \phantom{-}.037$^{*}$ & \phantom{-}.033$^{*}$ & \phantom{-}.025\phantom{$^{*}$} & -.020\phantom{$^{*}$} \\
\midrule
% Omni models (bottom)
Gemma4-E4B & -.003\phantom{$^{*}$} & -.027\phantom{$^{*}$} & -.015\phantom{$^{*}$} & -.003\phantom{$^{*}$} & -.005\phantom{$^{*}$} & \phantom{-}.004\phantom{$^{*}$} & \phantom{-}\textbf{.012}\phantom{$^{*}$} & \phantom{-}.007\phantom{$^{*}$} & -.014\phantom{$^{*}$} & \phantom{-}.009\phantom{$^{*}$} \\
MiniCPM-o-4.5 & \phantom{-}\textbf{.058}$^{*}$ & \phantom{-}.019\phantom{$^{*}$} & \phantom{-}.016\phantom{$^{*}$} & \phantom{-}.019\phantom{$^{*}$} & \phantom{-}.040$^{*}$ & \phantom{-}.017\phantom{$^{*}$} & \phantom{-}.031$^{*}$ & -.005\phantom{$^{*}$} & \phantom{-}.031$^{*}$ & \phantom{-}.003\phantom{$^{*}$} \\
Qwen3-Omni & \phantom{-}.057$^{*}$ & \phantom{-}.029$^{*}$ & \phantom{-}.032$^{*}$ & \phantom{-}.039$^{*}$ & \phantom{-}\textbf{.066}$^{*}$ & \phantom{-}.020$^{*}$ & \phantom{-}.044$^{*}$ & \phantom{-}.011\phantom{$^{*}$} & \phantom{-}.034$^{*}$ & \phantom{-}.001\phantom{$^{*}$} \\
Gemini3.5-Flash & -.008\phantom{$^{*}$} & \phantom{-}.043$^{*}$ & \phantom{-}\textbf{.104}$^{*}$ & -.002\phantom{$^{*}$} & -.006\phantom{$^{*}$} & \phantom{-}.010\phantom{$^{*}$} & -.007\phantom{$^{*}$} & -.008\phantom{$^{*}$} & \phantom{-}.011\phantom{$^{*}$} & \phantom{-}.011\phantom{$^{*}$} \\
\bottomrule
\end{tabular}
\end{table*}
\begin{table}[t]
\centering
\caption{\textbf{Experiment~3:} how often the model matches \emph{bouba} to the rounded shape (\textbf{B}) and \emph{kiki} to the pointed shape (\textbf{K}), averaged over the ten best-parsed languages.
\textbf{Experiment~4:} model agreement with humans on per-shape scores (first-order, Pearson $r$ and Spearman $r_s$) and on the $90\times90$ dissimilarity matrix (second-order RSA $\rho$). An asterisk ($^{*}$) marks correlations significant at $p<.05$.}
\vspace{-2mm}
\label{tab:exp3_and_4_merged}
\resizebox{\columnwidth}{!}{%
\begin{tabular}{lccccc}
\toprule
\multirow{3}{*}{Model} & \multicolumn{2}{c}{Experiment~3} & \multicolumn{3}{c}{Experiment~4} \\
\cmidrule(lr){2-3}\cmidrule(lr){4-6}
 & \multirow{2}{*}{B} & \multirow{2}{*}{K} & \multicolumn{2}{c}{First-order} & Second-order \\
\cmidrule(lr){4-5}\cmidrule(lr){6-6}
 & & & Pearson $r$ & Spearman $r_s$ & RSA $\rho$ \\
\midrule
Human & 0.81 & 0.62 & .991\phantom{$^{*}$} & .981\phantom{$^{*}$} & .906\phantom{$^{*}$} \\
\midrule
% Omni models
Gemma4-E4B & 0.51 & 0.50 & .953$^{*}$ & .870$^{*}$ & .807$^{*}$ \\
MiniCPM-o-4.5 & 0.67 & 0.64 & .974$^{*}$ & .928$^{*}$ & .845$^{*}$ \\
Qwen3-Omni & 0.18 & 0.77 & .939$^{*}$ & .841$^{*}$ & .861$^{*}$ \\
Gemini3.5-Flash & 1.00 & 1.00 & .970$^{*}$ & .918$^{*}$ & .864$^{*}$ \\
\bottomrule
\end{tabular}%
}
\end{table}
 
\subsection{Sound symbolism is weak in audio}
\label{sec:audio}
 
We begin by asking whether SLMs judge sounds as rounded or pointed the way humans do (RQ\,1).
Experiments~1 and~2 probe this in the auditory channel: a two-alternative forced choice in Experiment~1, and a graded rating over the full pseudoword set in Experiment~2.
 
In Experiment~1 (forced choice; Table~\ref{tab:exp1}), a superficially human-like pattern of \emph{bouba}-rounded rates exceeding \emph{kiki}-pointed rates can be produced by two response artifacts rather than by sound symbolism: a \emph{response bias}, in which the model favors one answer option regardless of the stimulus, and an \emph{order effect}, in which the answer depends on the position of the options in the prompt.
MiniCPM-o-4.5 shows the former, calling \emph{bouba} rounded on $.90$ of trials but also \emph{kiki} rounded on $.84$: it answers \emph{rounded} almost regardless of the sound, so its high \emph{bouba}-rounded rate reflects a default preference, not the acoustics.
Qwen3-Omni shows the latter, answering \emph{rounded} on $.89$ of trials when \emph{rounded} is listed first but only $.45$ when it is listed second (Wilcoxon signed-rank, $p=.004$), tracking option position rather than the sound.
Because both artifacts reproduce the human-like pattern without attending to the acoustics, the forced choice alone cannot establish sound symbolism, motivating the graded rating that follows.
 
Experiment~2 (graded rating; Table~\ref{tab:exp2}) gives a finer and larger test. The gap is unambiguous: the best model, Qwen3-Omni, reaches only $r=.470$ against a human ceiling of $.843$, and three of the eight models show no significant correlation at all.
Because the human ratings are themselves reliable, this is a real shortfall rather than a ceiling effect, with even the strongest model recovering little more than half of the achievable agreement.
The same weakness appears in how the models place the words relative to one another: their dissimilarity matrices match the human one at best $\rho=.104$ against a ceiling of $.377$, with several models near zero.
The shortfall thus extends from individual words to the global structure of the auditory space.
 
The same ratings let us locate the source of this shortfall (RQ\,2): we test whether a model's auditory space is shaped by the acoustic cues that predict human judgments (Table~\ref{tab:exp2_acoustic}).
For humans, spectral tilt dominates ($\rho=.431$), with the other cues far behind.
No model reproduces this profile: every model's correlations are far smaller, and their strongest cues scatter across largely unrelated parameters.
Even the largest correlation across all models and cues, Gemini3.5-Flash on the speech envelope ($\rho=.104$), falls on a cue that is secondary for humans and remains a quarter of the human spectral-tilt value.
The models therefore encode little of the acoustic structure that underlies human sound symbolism, pointing to the auditory representation itself as a likely locus of the weakness.
Whether the rounded/pointed distinction is representable by these models at all, or fails only in audio, is explored next in visual and crossmodal experiments.
 
\subsection{The crossmodal link fails on audio}
\label{sec:xm}

We test the crossmodal link directly in Experiment~3, asking whether an omni model can match a heard sound to one of two shapes, the setup closest to the original human task (RQ\,3; Table~\ref{tab:exp3_and_4_merged}).
Gemini3.5-Flash matches perfectly: it sends \emph{bouba} to the rounded shape and \emph{kiki} to the pointed shape at ceiling rates across the ten analyzed languages.
Given its weak auditory alignment in Experiment~2 ($r=.392$; acoustic-cue RSA near zero), this success is unlikely to rest on human-like acoustic representations; a plausible route is recognizing the two well-known pseudowords themselves and matching through lexical or semantic knowledge rather than the acoustics.
The three open-weight omni models all fail.
Qwen3-Omni shows a \emph{shape bias}, choosing the pointed shape on $.80$ of trials regardless of the sound, so both \emph{bouba} and \emph{kiki} land on it.
Gemma4-E4B shows the most extreme \emph{order effect}, picking the first-listed shape on every trial ($1.00$; Wilcoxon $p=.002$), so its flat $.50$ congruence is an artifact of counterbalancing rather than chance-level matching.
MiniCPM-o-4.5 lands only modestly above chance ($.67$ for \emph{bouba}, $.64$ for \emph{kiki}), but this too is inflated by an order effect, as it picks the first-listed shape on $.70$ of trials (Wilcoxon $p=.02$).

A matching failure could originate on the visual side, so we check shape perception directly. In Experiment~4 (visual-only ablation; Table~\ref{tab:exp3_and_4_merged}), the four omni models rate the shapes with no sound and track the human means closely: per-shape correlations run from $r=.94$ to $.97$ against a human ceiling of $.99$, and the models also group the shapes by similarity as people do, with second-order RSA from $\rho=.81$ to $.86$ against a ceiling of $.91$.
These near-ceiling scores show that shape perception is intact, so the open models' breakdown is localized to the auditory side rather than the visual one (RQ\,4): their auditory judgments are weak (Experiments~1 and~2) while their shape judgments are not, and the logit-lens analysis that follows further characterizes this modality gap.

\subsection{Perceptual structure forms only in deep layers}
To investigate how SLMs align their internal representations with human cognition, we apply a logit lens to Qwen3-Omni, the open-weight model with the strongest human alignment in Experiment~2 (Table~\ref{tab:exp2}).
We evaluate 536 audio and 90 image stimuli, prompting the model to rate each from 1 to 7 on a ``rounded'' scale and a ``pointed'' scale.
As described in Section~\ref{sec:logitlens}, for each stimulus we read the post-softmax distribution over the digit tokens \texttt{1}--\texttt{7} at every layer.
We track two metrics using line charts: (i) the maximum softmax probability over the seven digit tokens across all pooled stimuli, which reveals when the model commits to a rating rather than merely favoring it weakly, and (ii) the argmax digit per layer, averaged within human-rating bands.
For the second metric, we divide the stimuli by mean human rating into three groups: a low band $[1, 3)$, a mid band $[3, 5)$, and a high band $[5, 7]$.
This grouping lets us infer how the model internally aligns its layer-wise predictions with human cognitive categories.

\begin{figure}[t]
  \centering
  \includegraphics[width=\linewidth]{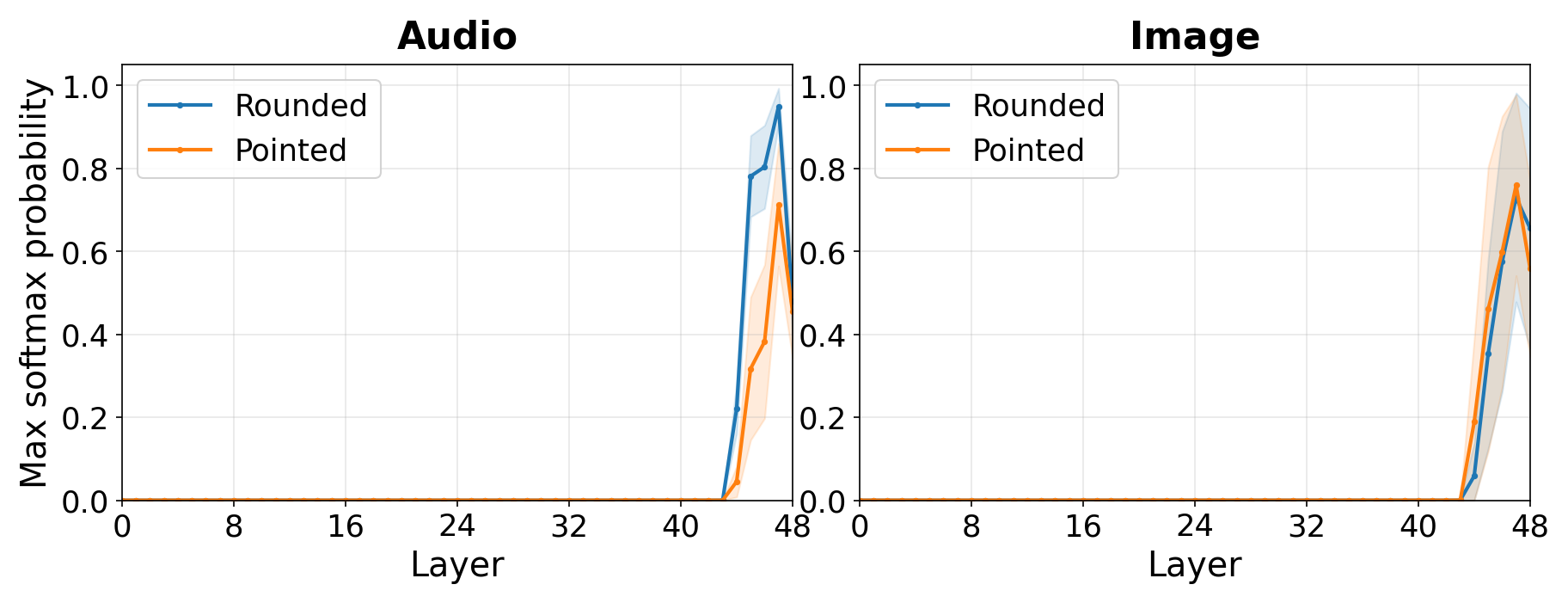}
  \vspace{-8mm}
  \caption{Layer-wise maximum digit-token probability for Qwen3-Omni on spoken pseudowords (Experiment~2, left) and abstract shapes (Experiment~4, right). At each layer, a full-vocabulary softmax is applied to the final token's residual stream. We plot the highest probability assigned to any digit token (\texttt{1}--\texttt{7}) per stimulus, averaged across all stimuli. Shaded bands show $\pm 1$ standard deviation across stimuli.}
  \vspace{-2mm}
  \label{fig:conf}
\end{figure}

In both modalities, Qwen3-Omni delays its commitments until the final 4 to 6 layers. As Fig.~\ref{fig:conf} demonstrates, the softmax distribution remains flat over the first $\sim\!90\%$ depth, indicating that early argmax trajectories lack meaningful partial answers. It is only in this final layer band, once a decision is actually formed, that the two modalities sharply diverge.

\begin{figure}[t]
  \centering
  \includegraphics[width=\linewidth]{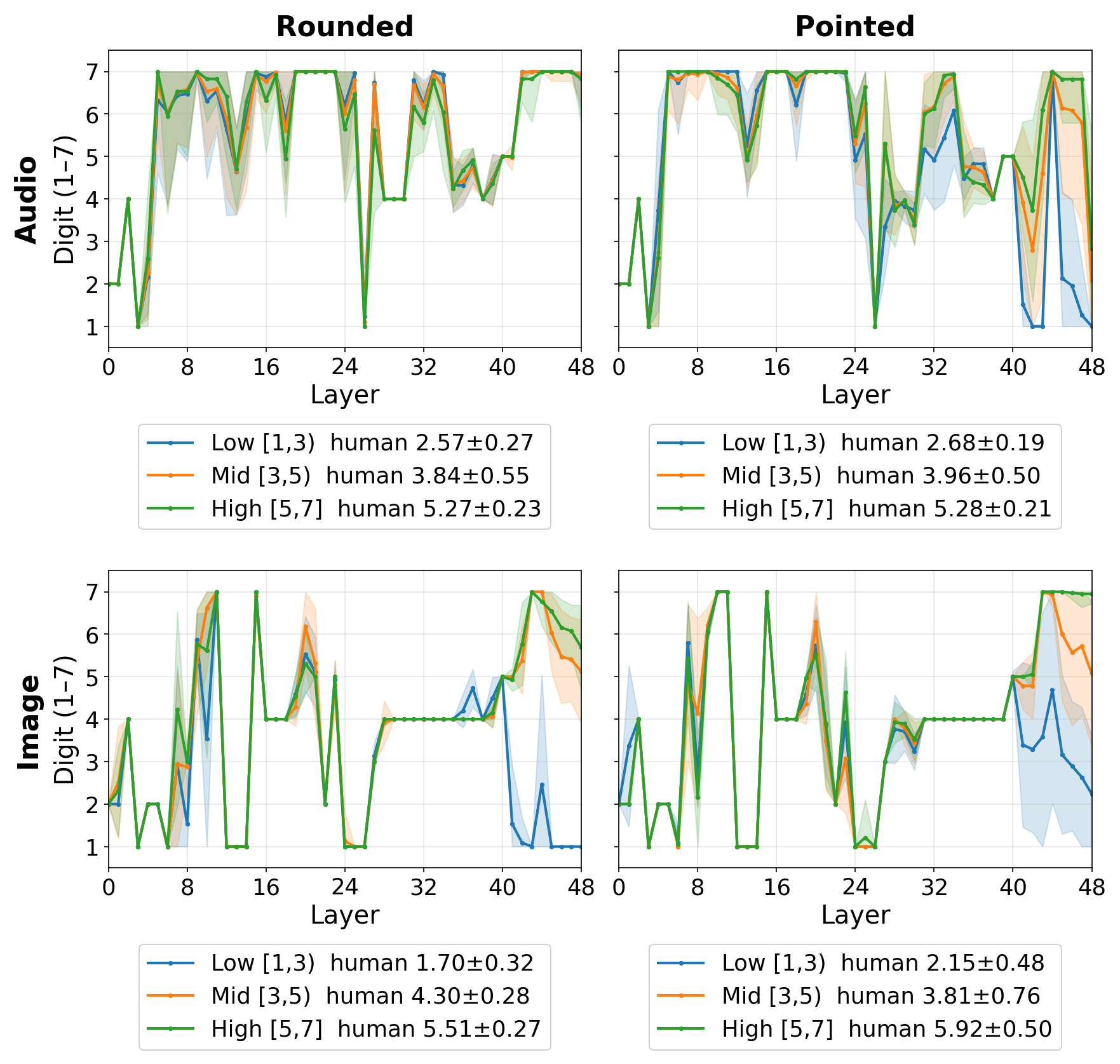}
  \vspace{-8mm}
  \caption{Logit-lens argmax-digit trajectories for Qwen3-Omni on spoken pseudowords (Experiment~2, top row) and abstract shapes (Experiment~4, bottom row) under ``rounded'' (left column) and ``pointed'' (right column) conditions. Stimuli are partitioned into three bands based on their \emph{mean human rating}. Annotations (human $\mu \pm \sigma$) report the distribution of these ratings, confirming the groups are perceptually distinct. Shaded bands represent $\pm 1$ standard deviation of the argmax digit across stimuli at each layer.}
  \vspace{-5mm}
  \label{fig:logit_lens}
\end{figure}

When processing spoken pseudowords in Fig.~\ref{fig:logit_lens}, the three human-rating bands stay almost entirely entangled across the full depth of Qwen3-Omni for both the rounded and the pointed settings.
The model converges to a shared 6--7 cluster for rounded with only slight separation at the last layer, and shows a faint late divergence for pointed where the low-rated band dips below the mid and high bands, indicating that the acoustic correlates of sound symbolism are not strongly reflected in its layer-wise digit predictions.
Conversely, when visualizing abstract shape images, Qwen3-Omni demonstrates a clear separation consistent with the human ratings.
This alignment is absent for most of the network but emerges abruptly late in the model from roughly the 40th layer, with the low band settling near digit 1 and the high band near digits 6--7.
Ultimately, these results highlight a modality gap; while the model lacks the strong ability to separate rounded and pointed in audio, it successfully captures these properties and aligns with the human ratings in the visual domain.

\section{Conclusion}
\label{sec:conclusion}
This work examined whether SLMs that directly process speech share the human link between sound and shape.
Separating the auditory, visual, and crossmodal sides of the effect reveals a clear dissociation.
These models read the visual structure of sound symbolism much as people do, yet their auditory judgments align only weakly with human ones and miss the acoustic cues that underlie human intuitions.
Because their shape perception stays intact, the breakdown localizes to how speech itself is represented.
The crossmodal link is achievable, as a frontier model succeeds at matching, though plausibly via lexical rather than acoustic routes, whereas the tested open-weight models do not.
Closing this gap will require audio representations that capture the acoustic cues humans rely on, a step toward models whose outputs match human sensory intuitions.

\bibliographystyle{IEEEtran}
\bibliography{IEEEref}

@article{cwiek2021bouba,
  title={The bouba/kiki effect is robust across cultures and writing systems},
  author={{\'C}wiek, Aleksandra and Fuchs, Susanne and Draxler, Christoph and Asu, Eva Liina and Dediu, Dan and Hiovain, Katri and Kawahara, Shigeto and Koutalidis, Sofia and Krifka, Manfred and Lippus, P{\"a}rtel and others},
  journal={Philosophical Transactions of the Royal Society B: Biological Sciences},
  volume={377},
  number={1841},
  pages={20200390},
  year={2021}
}

@article{lacey2020stimulus,
  title={Stimulus parameters underlying sound-symbolic mapping of auditory pseudowords to visual shapes},
  author={Lacey, Simon and Jamal, Yaseen and List, Sara M and McCormick, Kelly and Sathian, Krish and Nygaard, Lynne C},
  journal={Cognitive Science},
  volume={44},
  number={9},
  pages={e12883},
  year={2020},
  publisher={Wiley Online Library}
}

@inproceedings{mccormick2015,
  author    = {McCormick, Kelly and Kim, Jee Young and List, Sara and Nygaard, Lynne C.},
  title     = {Sound to meaning mappings in the bouba-kiki effect},
  booktitle = {Proceedings of the 37th Annual Conference of the Cognitive Science Society},
  pages     = {1565--1570},
  year      = {2015}
}

@misc{logit_lens,
  author = {nostalgebraist},
  title = {Interpreting {GPT}: the logit lens},
  howpublished = {\url{https://www.lesswrong.com/posts/AcKRB8wDpdaN6v6ru/interpreting-gpt-the-logit-lens}},
  year = {2020},
  month = {August},
  day = {31}
}

@article{ramachandran2001,
  author  = {Ramachandran, V. S. and Hubbard, E. M.},
  title   = {Synaesthesia --- A Window into Perception, Thought and Language},
  journal = {Journal of Consciousness Studies},
  year    = {2001},
  volume  = {8},
  number  = {12},
  pages   = {3--34},
}

@article{dingemanse2015arbitrariness,
  title={Arbitrariness, iconicity, and systematicity in language},
  author={Dingemanse, Mark and Blasi, Dami{\'a}n E and Lupyan, Gary and Christiansen, Morten H and Monaghan, Padraic},
  journal={Trends in cognitive sciences},
  volume={19},
  number={10},
  pages={603--615},
  year={2015},
  publisher={Elsevier}
}

@book{hinton1994sound,
  title={Sound symbolism},
  author={Hinton, Leanne and Nichols, Johanna and Ohala, John J},
  year={1994},
  publisher={Cambridge University Press}
}

@article{ohala1984,
  author  = {Ohala, John J.},
  title   = {An Ethological Perspective on Common Cross-Language Utilization of {F0} of Voice},
  journal = {Phonetica},
  year    = {1984},
  volume  = {41},
  pages   = {1--16},
  doi     = {10.1159/000261706},
}

@inproceedings{loakman2024,
  author    = {Loakman, Tyler and Li, Yucheng and Lin, Chenghua},
  title     = {With Ears to See and Eyes to Hear: Sound Symbolism Experiments with Multimodal Large Language Models},
  booktitle = {Proceedings of the 2024 Conference on Empirical Methods in Natural Language Processing (EMNLP)},
  year      = {2024},
  pages     = {2849--2867},
  doi       = {10.18653/v1/2024.emnlp-main.167},
}

@inproceedings{jeong2026language,
  title={Do Language Models Associate Sound with Meaning? A Multimodal Study of Sound Symbolism},
  author={Jeong, Jinhong and Lee, Sunghyun and Lee, Jaeyoung and Han, Seonah and Yu, Youngjae},
  booktitle={Proceedings of the AAAI Conference on Artificial Intelligence},
  volume={40},
  number={37},
  pages={31247--31255},
  year={2026}
}

@inproceedings{iida2024investigating,
  title={Investigating Iconicity in Vision-and-Language Models: A Case Study of the Bouba/Kiki Effect in Japanese Models},
  author={Iida, Hinano and Funakura, Hayate},
  booktitle={Proceedings of the Annual Meeting of the Cognitive Science Society},
  volume={46},
  year={2024}
}

@article{alper2023kiki,
  title={Kiki or bouba? sound symbolism in vision-and-language models},
  author={Alper, Morris and Averbuch-Elor, Hadar},
  journal={Advances in Neural Information Processing Systems},
  volume={36},
  pages={78347--78359},
  year={2023}
}

@article{xu2025qwen3,
  title={Qwen3-omni technical report},
  author={Xu, Jin and Guo, Zhifang and Hu, Hangrui and Chu, Yunfei and Wang, Xiong and He, Jinzheng and Wang, Yuxuan and Shi, Xian and He, Ting and Zhu, Xinfa and others},
  journal={arXiv preprint arXiv:2509.17765},
  year={2025}
}

@article{cui2026minicpm,
  title={Minicpm-o 4.5: Towards real-time full-duplex omni-modal interaction},
  author={Cui, Junbo and Xu, Bokai and Wang, Chongyi and Yu, Tianyu and Sun, Weiyue and Xu, Yingjing and Wang, Tianran and He, Zhihui and Ma, Wenshuo and Cai, Tianchi and others},
  journal={arXiv preprint arXiv:2604.27393},
  year={2026}
}

@article{wu2025step,
  title={Step-audio 2 technical report},
  author={Wu, Boyong and Yan, Chao and Hu, Chen and Yi, Cheng and Feng, Chengli and Tian, Fei and Shen, Feiyu and Yu, Gang and Zhang, Haoyang and Li, Jingbei and others},
  journal={arXiv preprint arXiv:2507.16632},
  year={2025}
}

@article{ding2025kimi,
  title={Kimi-audio technical report},
  author={Ding, Ding and Ju, Zeqian and Leng, Yichong and Liu, Songxiang and Liu, Tong and Shang, Zeyu and Shen, Kai and Song, Wei and Tan, Xu and Tang, Heyi and others},
  journal={arXiv preprint arXiv:2504.18425},
  year={2025}
}

@inproceedings{NEURIPS2025_3babb6b4,
 author = {Ghosh, Sreyan and Goel, Arushi and Kim, Jaehyeon and Kumar, Sonal and Kong, Zhifeng and Lee, Sang-gil and Yang, Chao-Han and Duraiswami, Ramani and Manocha, Dinesh and Valle, Rafael and Catanzaro, Bryan},
 booktitle = {Advances in Neural Information Processing Systems},
 editor = {D. Belgrave and C. Zhang and H. Lin and R. Pascanu and P. Koniusz and M. Ghassemi and N. Chen},
 pages = {41819--41886},
 publisher = {Curran Associates, Inc.},
 title = {Audio Flamingo 3: Advancing Audio Intelligence with Fully Open Large Audio Language Models},
 url = {https://proceedings.neurips.cc/paper_files/paper/2025/file/3babb6b453cb59d87cb58a1219ef914b-Paper-Conference.pdf},
 volume = {38},
 year = {2025}
}

@article{mantel1967detection,
  title={The detection of disease clustering and a generalized regression approach},
  author={Mantel, Nathan},
  journal={Cancer research},
  volume={27},
  number={2\_Part\_1},
  pages={209--220},
  year={1967},
  publisher={The American Association for Cancer Research}
}

@article{spearman1910correlation,
  title={Correlation calculated from faulty data},
  author={Spearman, Charles},
  journal={British journal of psychology},
  volume={3},
  number={3},
  pages={271},
  year={1910},
  publisher={Cambridge University Press}
}

@article{brown1910some,
  title={Some experimental results in the correlation of mental abilities 1},
  author={Brown, William},
  journal={British Journal of Psychology, 1904-1920},
  volume={3},
  number={3},
  pages={296--322},
  year={1910},
  publisher={Blackwell Publishing Ltd Oxford, UK}
}

@article{sapir1929study,
  title={A study in phonetic symbolism.},
  author={Sapir, Edward},
  journal={Journal of experimental psychology},
  volume={12},
  number={3},
  pages={225},
  year={1929},
  publisher={Psychological Review Company}
}

@article{kohler1967gestalt,
  title={Gestalt psychology},
  author={K{\"o}hler, Wolfgang},
  journal={Psychologische forschung},
  volume={31},
  number={1},
  pages={XVIII--XXX},
  year={1967},
  publisher={Springer}
}

@inproceedings{shinto2024analyzing,
  title={Analyzing the Sensibility of Visual Language Models Using an Evolving Image Generation System: Focusing on Color Impressions and Sound Symbolism},
  author={Shinto, Ryoma and Iizuka, Hiroyuki},
  booktitle={Artificial Life Conference Proceedings},
  volume={36},
  year={2024},
  doi={10.1162/isal_a_00719},
}

@inproceedings{dynamicsuperb2,
title={Dynamic-{SUPERB} Phase-2: A Collaboratively Expanding Benchmark for Measuring the Capabilities of Spoken Language Models with 180 Tasks},
author={Huang, Chien-yu and others},
booktitle={The Thirteenth International Conference on Learning Representations},
year={2025}
}

@inproceedings{wang-etal-2025-audiobench,
    title = "{A}udio{B}ench: A Universal Benchmark for Audio Large Language Models",
    author = "Wang, Bin  and
      Zou, Xunlong  and
      Lin, Geyu  and
      Sun, Shuo  and
      Liu, Zhuohan  and
      Zhang, Wenyu  and
      Liu, Zhengyuan  and
      Aw, AiTi  and
      Chen, Nancy F.",
    editor = "Chiruzzo, Luis  and
      Ritter, Alan  and
      Wang, Lu",
    booktitle = "Proceedings of the 2025 Conference of the Nations of the Americas Chapter of the Association for Computational Linguistics: Human Language Technologies (Volume 1: Long Papers)",
    month = apr,
    year = "2025",
    address = "Albuquerque, New Mexico",
    publisher = "Association for Computational Linguistics",
    url = "https://aclanthology.org/2025.naacl-long.218/",
    doi = "10.18653/v1/2025.naacl-long.218",
    pages = "4297--4316",
    ISBN = "979-8-89176-189-6"
}

@article{mmaupro,
  title={Mmau-pro: A challenging and comprehensive benchmark for holistic evaluation of audio general intelligence},
  author={Kumar, Sonal and Sedl{\'a}{\v{c}}ek, {\v{S}}imon and Lokegaonkar, Vaibhavi and L{\'o}pez, Fernando and Yu, Wenyi and Anand, Nishit and Ryu, Hyeonggon and Chen, Lichang and Pli{\v{c}}ka, Maxim and Hlav{\'a}{\v{c}}ek, Miroslav and others},
  journal={arXiv preprint arXiv:2508.13992},
  year={2025}
}

@inproceedings{speechifeval,
  title     = {{Speech-IFEval: Evaluating Instruction-Following and Quantifying Catastrophic Forgetting in Speech-Aware Language Models}},
  author    = {Ke-Han Lu and others},
  year      = {2025},
  booktitle = {{Interspeech 2025}},
  pages     = {2078--2082},
  doi       = {10.21437/Interspeech.2025-619},
  issn      = {2958-1796},
}

@article{zhao2026evidence,
  title={Evidence for systematic semantic structure in individual phonemes},
  author={Zhao, Gexin},
  journal={arXiv preprint arXiv:2603.17306},
  year={2026}
}

@misc{sakshi2024mmaumassivemultitaskaudio,
      title={MMAU: A Massive Multi-Task Audio Understanding and Reasoning Benchmark}, 
      author={S Sakshi and Utkarsh Tyagi and Sonal Kumar and Ashish Seth and Ramaneswaran Selvakumar and Oriol Nieto and Ramani Duraiswami and Sreyan Ghosh and Dinesh Manocha},
      year={2024},
      eprint={2410.19168},
      archivePrefix={arXiv},
      primaryClass={eess.AS},
      url={https://arxiv.org/abs/2410.19168}, 
}

@inproceedings{listentthinkunderstand,
 author = {Gong, Yuan and Luo, Hongyin and Liu, Alexander and Karlinsky, Leonid and Glass, James R},
 booktitle = {International Conference on Learning Representations},
 editor = {B. Kim and Y. Yue and S. Chaudhuri and K. Fragkiadaki and M. Khan and Y. Sun},
 pages = {18516--18545},
 title = {Listen, Think, and Understand},
 url = {https://proceedings.iclr.cc/paper_files/paper/2024/file/510d0935b543a29d686f93fa52d1c288-Paper-Conference.pdf},
 volume = {2024},
 year = {2024}
}

@article{zhou2025echomind,
  title={EchoMind: An Interrelated Multi-level Benchmark for Evaluating Empathetic Speech Language Models},
  author={Zhou, Li and Yu, Lutong and Lyu, You and Lin, Yihang and Zhao, Zefeng and Ao, Junyi and Zhang, Yuhao and Wang, Benyou and Li, Haizhou},
  journal={arXiv preprint arXiv:2510.22758},
  year={2025}
}

@article{wang2021generating,
  title={Generating images from spoken descriptions},
  author={Wang, Xinsheng and Qiao, Tingting and Zhu, Jihua and Hanjalic, Alan and Scharenborg, Odette},
  journal={IEEE/ACM Transactions on Audio, Speech, and Language Processing},
  volume={29},
  pages={850--865},
  year={2021},
  publisher={IEEE}
}

@inproceedings{yang-etal-2025-towards-holistic,
    title = "Towards Holistic Evaluation of Large Audio-Language Models: A Comprehensive Survey",
    author = "Yang, Chih-Kai  and
      Ho, Neo S.  and
      Lee, Hung-yi",
    editor = "Christodoulopoulos, Christos  and
      Chakraborty, Tanmoy  and
      Rose, Carolyn  and
      Peng, Violet",
    booktitle = "Proceedings of the 2025 Conference on Empirical Methods in Natural Language Processing",
    month = nov,
    year = "2025",
    address = "Suzhou, China",
    publisher = "Association for Computational Linguistics",
    url = "https://aclanthology.org/2025.emnlp-main.514/",
    doi = "10.18653/v1/2025.emnlp-main.514",
    pages = "10144--10170",
    ISBN = "979-8-89176-332-6"
}

@INPROCEEDINGS{dynamicsuperb1,
  author={Huang, Chien-Yu and Lu, Ke-Han and Wang, Shih-Heng and Hsiao, Chi-Yuan and Kuan, Chun-Yi and Wu, Haibin and Arora, Siddhant and Chang, Kai-Wei and Shi, Jiatong and Peng, Yifan and Sharma, Roshan and Watanabe, Shinji and Ramakrishnan, Bhiksha and Shehata, Shady and Lee, Hung-Yi},
  booktitle={ICASSP 2024 - 2024 IEEE International Conference on Acoustics, Speech and Signal Processing (ICASSP)}, 
  title={Dynamic-Superb: Towards a Dynamic, Collaborative, and Comprehensive Instruction-Tuning Benchmark For Speech}, 
  year={2024},
  volume={},
  number={},
  pages={12136-12140},
  doi={10.1109/ICASSP48485.2024.10448257}}

@article{chen2026voicebench,
  title={Voicebench: Benchmarking llm-based voice assistants},
  author={Chen, Yiming and Yue, Xianghu and Zhang, Chen and Gao, Xiaoxue and Tan, Robby T and Li, Haizhou},
  journal={Transactions of the Association for Computational Linguistics},
  volume={14},
  pages={378--398},
  year={2026},
  publisher={MIT Press 255 Main Street, 9th Floor, Cambridge, Massachusetts 02142, USA~…}
}

@misc{google2025gemma4,
  title        = {Gemma 4 Model Card},
  author       = {{Google DeepMind}},
  year         = {2026},
  howpublished = {\url{https://ai.google.dev/gemma/docs/core/model_card_4}},
}

\end{document}